\documentclass[twocolumn,showpacs,preprintnumbers,amsmath,amssymb,prb]{revtex4}

\usepackage{graphicx}% Include figure files
\usepackage{dcolumn}% Align table columns on decimal point
\usepackage{bm}% bold math

\def\U#1{{%
\def\O{\mbox{O}}
\def\u{\mbox{u}}
\mathcode`\u=\mu
\mathcode`\O=\Omega
\mathrm{#1}}}

\def\dd{{\mathrm{d}}}
\def\sub#1{_{\scriptsize\mbox{#1}}}

\def\dquote#1{\textquotedblleft #1\textquotedblright}
\def\degree{\mbox{$^\circ$}}
\def\vct#1{{\mathchoice{\mbox{\boldmath$#1$}}{\mbox{\boldmath$#1$}}%
  {\mbox{\scriptsize\boldmath$#1$}}{\mbox{\scriptsize\boldmath$#1$}}}}

\begin{document}

%\preprint{}

\title{Variable group delay in a
metamaterial with field-gradient-induced transparency}
% Force line breaks with \\

\author{Yasuhiro Tamayama}
\email{tamayama.yasuhiro.83a@st.kyoto-u.ac.jp}
\author{Toshihiro Nakanishi}
\author{Masao Kitano}
 \affiliation{Department of Electronic Science and
 Engineering, Kyoto University, Kyoto 615-8510, Japan}

\date{\today}% It is always \today, today,

\begin{abstract}

We realize variable control of the group delay in an
electromagnetically induced transparency-like metamaterial.
Its unit cell is designed to have a bright mode and a dark mode.
The coupling strength between these two modes is determined by the
electromagnetic field gradient. 
In this metamaterial with field-gradient-induced transparency,
the group delay at the transparency frequency can be varied by
varying the incident angle of the electromagnetic plane waves.
By tilting a single layer of the metamaterial, the group delay of a
microwave pulse can be varied between 0.50 and 1.85 ns.

\end{abstract}

\maketitle

Electromagnetically induced transparency (EIT) has attracted considerable
attention in recent years as a means to control the group
velocity
of electromagnetic waves.\cite{harris97,hau99,fleischhauer05} 
It is a quantum interference phenomenon that
arises in $\Lambda$-type three-state atoms that
interact with two electromagnetic waves: a probe wave, which is tuned to the
transition between the ground state and the 
common excited state of the atom, and a coupling wave, which
is tuned to the transition between the 
intermediate and excited states. 
The atoms absorb the probe wave in the absence
of the coupling wave, but this absorption is
suppressed in a narrow frequency range when a coupling wave is present. 
From the Kramers-Kronig relations,\cite{saleh07} we see that the narrow
dip of absorption associates 
a steep change in dispersion (refractive index) for the probe wave. 
The steep dispersion change results in the slow group velocity of
the probe wave because the group velocity is given by $v\sub{g} = c_0 /
[n+\omega(\dd n / \dd \omega)]$, where $n (\omega)$ is
the refractive index, $\omega$ is the angular frequency, and $c_0$ is
the speed of light in vacuum. Since the bandwidth of the transmission
window, which determines the steepness of the dispersion, depends on the 
Rabi frequency of the coupling wave, the group velocity can be controlled by
varying the intensity of the coupling wave. 

Since rather complex experimental setups are required to
generate EIT, several studies have sought to mimic 
the effect in classical
systems,\cite{smith04,yanik04_2,totsuka07,yang09,liu_opex09,fedotov07,zhang_prl08,liu_am08,papasimakis08,papasimakis09,tassin_prl09,tassin_opex09,liu_nat09,yannopapas09,chiam09,verellen09,tamayama10,lu10_opex,kurter11} 
especially in 
metamaterials.\cite{fedotov07,zhang_prl08,liu_am08,papasimakis08,papasimakis09,tassin_prl09,tassin_opex09,liu_nat09,yannopapas09,chiam09,verellen09,tamayama10,lu10_opex,kurter11} 
Metamaterials used to mimic EIT are often
designed based on the classical analog of EIT.\cite{alzar02} 
The unit cell of EIT-like metamaterials
consists of two coupled resonant modes: a
low-quality-factor ($Q$) resonant mode (bright mode) 
that can be directly excited by
the incident wave and a high-$Q$ resonant mode (dark mode) 
that cannot be directly excited by the incident wave. 

Variable group delay in EIT-like
metamaterials, as in the original EIT, is potentially useful for 
controlling electromagnetic pulse propagation. 
The coupling strength between the bright and dark modes
needs to be controlled to vary the group delay. 
In early studies on EIT-like
metamaterials,\cite{fedotov07,zhang_prl08,liu_am08,tassin_prl09,tassin_opex09,liu_nat09,verellen09}
the coupling strength was controlled by adjusting 
the geometrical parameters of metamaterials; 
however, this method cannot be practically applied to achieve variable
control of the group delay. 
Later, we\cite{tamayama10} and another group\cite{lu10_opex} 
independently proposed a method for achieving variable control of the group
delay without changing the geometrical 
parameters. However, this method has not been experimentally verified.
In this paper, we demonstrate variable control of the group
delay in a metamaterial with field-gradient-induced transparency,
which is the
EIT-like metamaterial developed in our previous study,\cite{tamayama10}
by performing time-domain measurements.

\begin{figure}[tb]
\begin{center}
\includegraphics[scale=0.5]{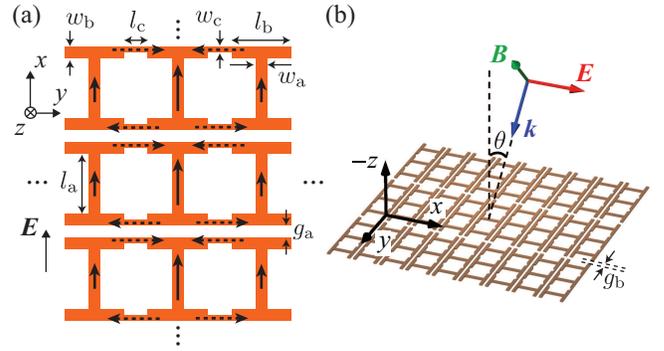}
\caption{(Color online) (a) Schematic of
a metallic metamaterial with field-gradient-induced transparency. Every
 three I-shaped units are connected in the $y$ direction. The solid
 (dashed) arrows indicate the current flow of the bright (dark)
 mode. (b) Orientation of the metamaterial with respect to 
 the incident plane wave (transverse-electric wave). 
 The magnetic field $\vct{B}$ and 
 wavevector $\vct{k}$ are in the $yz$ plane
 and the electric field $\vct{E}$ is in the $x$ direction. }
\label{fig:1}
\end{center}
\end{figure}

Figure \ref{fig:1}(a) shows a schematic of 
the metamaterial used in this study. We
focus on two resonant modes in this metamaterial:
electric dipole resonance (the solid arrows indicate its current flow)
and magnetic quadrupole-like
resonance (represented by the dashed arrows);
these modes are, respectively, referred to as modes 1 and 2. 
Mode 2 has a higher $Q$ value than mode 1 because the magnetic 
quadrupole-like 
oscillator emits less electromagnetic waves than the electric dipole
oscillator.

The $x$-polarized incident electric field can directly excite only mode 1. 
Mode 2 cannot be directly excited because the current
flow of mode 2 is perpendicular to the incident electric field. 
The two resonant modes are magnetically coupled
when the incident electromagnetic field has a field gradient. 
When the $x$-polarized electric field has a 
gradient in the $y$ direction, the induced current of mode 1 
depends on $y$. The difference between adjacent currents generates a
magnetic flux in the loop of mode 2 and induces antiparallel currents
via the electromotive force. 
Mode 1, mode 2, and the field gradient in the $y$ direction 
of the $x$ component of the incident electric field
correspond to the bright mode, dark mode, 
and coupling strength between these two modes, respectively. 
Therefore, this system is similar to the classical model
of EIT\cite{alzar02} and the metamaterial behaves as an EIT-like medium
for the $x$-polarized electromagnetic wave. 

To achieve variable control of the group delay in the EIT-like 
metamaterial, the coupling strength between the two resonant modes 
needs to be controlled. The coupling strength depends on 
the gradient of the $x$ component of the incident electric field in the
$y$ direction. Thus, the 
group delay can be varied by changing, for example, the beamwidths of
normally incident Gaussian beams, the widths of
waveguides,\cite{tamayama10} or the incident angle $\theta$ 
of the plane wave defined in Fig.\,\ref{fig:1}(b). 
In the present study, 
we investigate the $\theta$ dependence of the transmission properties
of the metamaterial.
(One might be concerned that the direct interaction between the incident
magnetic field and the dark mode occurs when $\theta$ is large, which
causes a reduction in the transmittance at the transparency
frequency. However, this effect 
is negligible for the magnetic quadrupole-like dark mode.)

We fabricated the metamaterial shown in Fig.\,\ref{fig:1} using a printed
circuit board that consisted of a $35$-$\U{um}$-thick copper film on a
$0.8$-$\U{mm}$-thick polyphenylene ether substrate with a relative
permittivity of $3.3$ and a loss tangent of $0.005$ at $6\,\U{GHz}$. 
The geometrical parameters defined in Fig.\,\ref{fig:1} 
are $l\sub{a}= 7.8\,\U{mm}$, $l\sub{b}=
11.0\,\U{mm}$, $l\sub{c}= 1.2\,\U{mm}$, $w\sub{a}= 5.0\,\U{mm}$, $w\sub{b}=
2.0\,\U{mm}$, $w\sub{c}= 1.0\,\U{mm}$, $g\sub{a}= 0.4\,\U{mm}$, and 
$g\sub{b}=1.2\,\U{mm}$. 

We measured transmission spectra of the fabricated metamaterial for
different incident angles to confirm whether an EIT-like transparency
phenomenon is observed.
A layer of the metamaterial was placed in free space and two horn
antennas connected to a network analyzer were used as a microwave
transmitter and receiver.

\begin{figure}[tb]
\begin{center}
\includegraphics[scale=1]{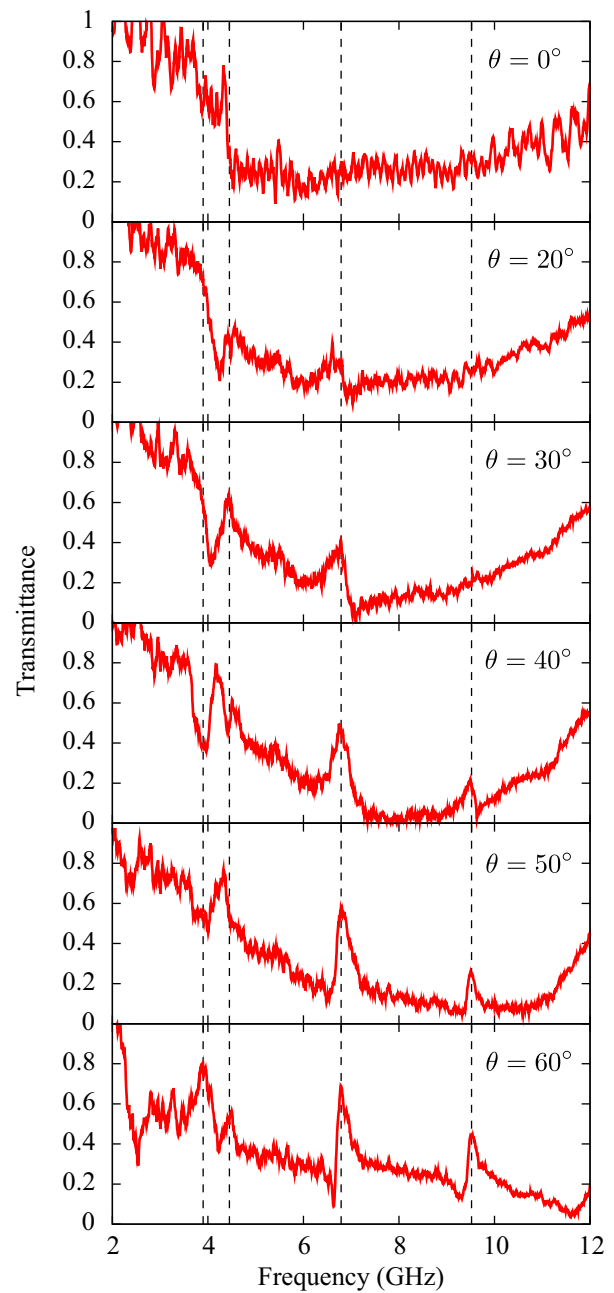}
\caption{(Color online) Transmission spectra of the fabricated
 metamaterial for six different incident angles $\theta$. The vertical
 dashed lines serve as guides to the eye for comparing the transmission
 peak frequencies for different $\theta$.}
\label{fig:2}
\end{center}
\end{figure}

\begin{figure}[tb]
\begin{center}
\includegraphics[scale=0.8]{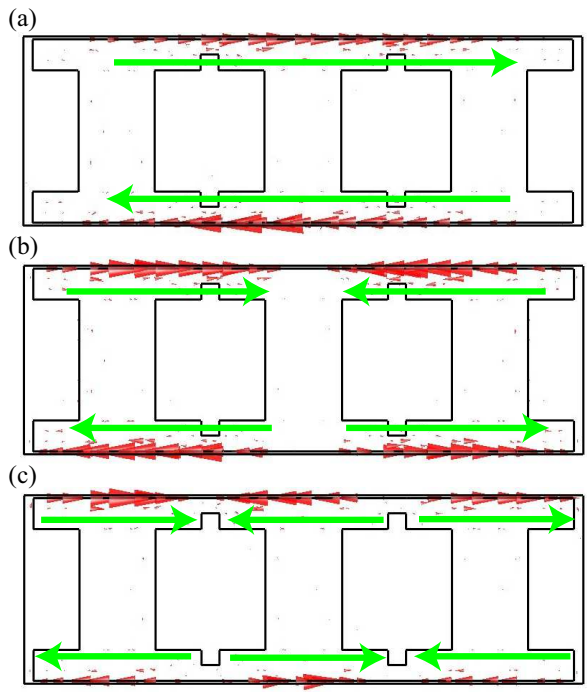}
\caption{(Color online) Current distributions 
 at the (a) lowest, (b) middle, and (c) highest
 transmission peak frequencies. The size 
 of the red (dark gray) cones represents the
 current magnitude. The green (light gray) arrows are guides to the
 eye and represent the direction of the current flow. 
The perfect electric conductor boundary
 conditions are applied to the vertical direction and the Floquet
 boundary conditions are applied to the horizontal direction. }
\label{fig:2_2}
\end{center}
\end{figure}

Figure \ref{fig:2} shows the transmission spectra of the
metamaterial obtained at six different incident angles. 
Only a broad absorption line centered on $6.0\,\U{GHz}$
(the resonant frequency of the bright mode)
is observed for normal incidence, whereas
three narrow transmission windows appear at about 4.2,
6.79, and $9.52\,\U{GHz}$ in the absorption line for oblique
incidence. This implies that an EIT-like transparency phenomenon 
occurs at these three frequencies. 
The transmission peak at about $4.2\,\U{GHz}$ shifts to lower
frequencies with increasing the incident angle, whereas the peaks at 6.79
and $9.52\,\U{GHz}$ exhibit little frequency shift. The transmittance at
$6.79\,\U{GHz}$ is higher than that at $9.52\,\U{GHz}$. Therefore, 
we consider the transparency window at $6.79\,\U{GHz}$ to be suitable
for controlling the group delay. 

In order to clarify the resonant modes at the
three transmission peak frequencies, we analyzed the
metamaterial using a finite element solver, COMSOL Multiphysics. 
The geometrical parameters used in the numerical 
analysis were the same as that in
the experiment except for the thickness of the copper layer. We set the
thickness of the copper layer to $500\,\U{um}$ for reducing
memory consumption in the simulation. This change causes no substantial
influence on the simulation results. 

Three transmission windows in a broad absorption line for
oblique incidence were also observed in the simulation. 
The current distributions at the frequencies
corresponding to the three transmission peaks are shown in 
Fig.\,\ref{fig:2_2}. Note that no significant currents are induced in the
bright mode at these frequencies. The lowest, middle, and highest
frequency resonant modes are found to be a magnetic
dipole resonance, an antisymmetric magnetic dipole resonance (magnetic
quadrupole-like resonance), and an incompletely antisymmetric magnetic
dipole resonance, respectively. It is the middle frequency mode
that we described in Fig.\,\ref{fig:1}(a) as the dark
mode. The middle frequency mode, 
or the magnetic quadrupole-like oscillator, 
has the highest $Q$ value in the
three resonant modes because the current distribution is most
antisymmetric. This characteristic causes the high transmittance 
and small frequency shift of the peak. We find also from
the current distribution that 
the transparency window corresponding to the middle frequency mode is 
most suitable. 

\begin{figure}[tb]
\begin{center}
\includegraphics[scale=0.7]{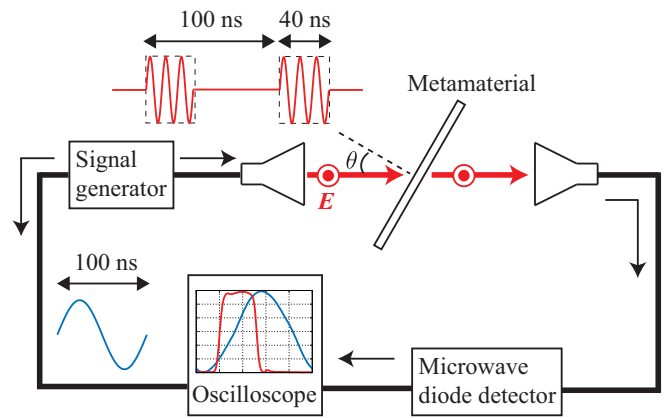}
\caption{(Color online) Schematic of experimental setup for 
pulse transmission measurements. }
\label{fig:3}
\end{center}
\end{figure}

\begin{figure}[tb]
\begin{center}
\includegraphics[scale=1]{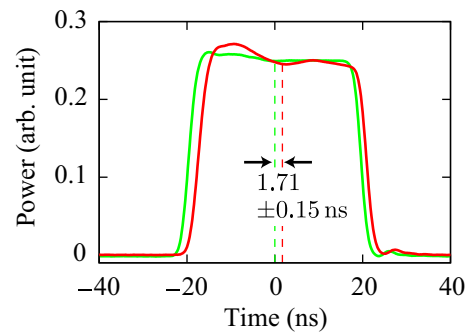}
\caption{(Color online) Envelope of the transmitted pulse for 
 $\theta = 40\degree$ [red (dark gray)] and that obtained 
 without the metamaterial
 [green (light gray)]. These wave forms are both averages obtained for $1000$
 pulses. The heights of the wave forms have been normalized. }
\label{fig:4}
\end{center}
\end{figure}

We performed pulse transmission 
measurements to evaluate the group delay in the
metamaterial. Figure \ref{fig:3} shows a schematic of the experimental
setup. The arrangement of the
metamaterial and the horn antennas was the same as that used to measure 
the transmission spectrum. A
signal generator was used to generate a microwave pulse
with a carrier frequency of $6.79\,\U{GHz}$, a pulse width of
$40\,\U{ns}$, and a period of $100\,\U{ns}$.
This pulse was emitted from the
horn antenna and was incident on the single layer of the metamaterial. 
The transmitted wave was received by the other horn antenna and 
detected by a microwave diode detector. The output
signal of the diode detector (i.e., the envelope of the transmitted pulse)
was observed using an oscilloscope. 
A $10$-$\U{MHz}$ sinusoidal wave generated by the signal generator was
used as the trigger signal for the oscilloscope. 
The transmittance and group delay
of the metamaterial were determined by comparing the transmitted pulses
obtained with and without the metamaterial.

Figure \ref{fig:4} shows the envelope of the transmitted pulse
for $\theta = 40\degree$ and that obtained without
the metamaterial. The former pulse is delayed by $1.71\,\U{ns}$ relative to
the latter pulse. We regard the delay of the
pulse center as being the group delay because it can be shown
analytically for Gaussian pulses\cite{yariv97} and numerically for other
pulses that these delays coincide if 
third- and higher-order dispersion are negligible. 
In fact, third- and higher-order dispersion are considered to be small
since the transmitted pulse does not exhibit any significant
distortion. 
Note that the width of the transmitted pulse varies due to
second-order dispersion (i.e., group delay dispersion).

\begin{figure}[tb]
\begin{center}
\includegraphics[scale=1]{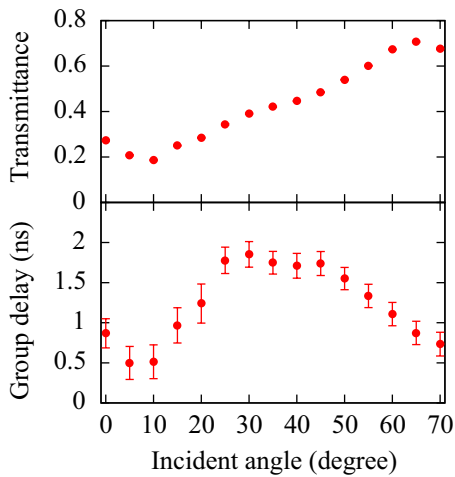}
\caption{(Color online) Measured transmittance and group delay as a function of
 incident angle for a pulse with a carrier frequency of
 $6.79\,\U{GHz}$. }
\label{fig:5}
\end{center}
\end{figure}

Figure \ref{fig:5} shows the measured transmittance
and group delay as a function of the incident angle. 
The transmittance decreases with decreasing the incident angle, or the
coupling strength between the bright and dark modes, due to the finite
loss in the dark mode. As the incident angle decreases, the group delay
initially increases and reaches a maximum at about $30\degree$, where
the transmission bandwidth becomes minimum and the steepest dispersion
is obtained. The incident angle for which the group delay takes a maximum
value is determined by the losses in the bright and dark
modes.\cite{zhang_prl08,tamayama10} When the incident angle becomes
smaller than $30\degree$, the transparency window gradually disappears  
and the group delay decreases. 
The group delay is found to be tunable in the range 0.50--$1.85\,\U{ns}$. 
This observation is consistent with the characteristics of
EIT, in which the transmittance decreases with
decreasing the intensity of the coupling wave and the group delay
takes a maximum value at a certain intensity of the coupling 
wave.\cite{fleischhauer05} 

In conclusion, we have achieved variable control of the group delay in a
metamaterial with field-gradient-induced transparency. 
The metamaterial consists of bright and dark modes that are coupled
with each other. The group delay in the metamaterial depends on the
coupling strength between these two modes. The coupling strength
is determined by the field gradient of the incident electromagnetic
wave, which allows the group delay to be varied by changing 
the incident angle.
We examined the incident angle dependence of the transmission spectrum
of the metamaterial. A broad absorption spectrum was obtained for
normal incidence, whereas EIT-like narrow transmission windows were observed for
oblique incidence. Pulse measurements revealed that the group
delay varies with the incident angle. 
The maximum group delay
can be increased if the $Q$ value of the dark mode is increased. 
The small deformation of the transmitted pulse observed in the present
experiment can be further suppressed
by matching the resonant frequency of the bright mode with that of the
dark mode because the shape of the transmission window becomes more
symmetric. 

This research was supported in part by Grants-in-Aid for Scientific Research 
(Grants No.\,22109004 and No.\,22560041), by the Global COE Program 
\dquote{Photonics and Electronics 
Science and Engineering} at Kyoto University, and by a research grant
from the Murata Science Foundation. 
One of the authors (Y.T.) acknowledges support 
from the Japan Society for the Promotion of Science.

%\newpage %Just because of unusual number of tables stacked at end
%\bibliography{main}% Produces the bibliography via BibTeX.

%

\end{document}